\documentclass[aps,twocolumn,showpacs, showkeys,nofootinbib,floatfix]{revtex4-1}

\usepackage{amsmath}
\usepackage{amsfonts}
\usepackage{txfonts}
\usepackage{graphicx}
\usepackage{dcolumn}
\usepackage{bbm}
\usepackage{amssymb}
\usepackage{latexsym}
\usepackage{titlesec}
\usepackage[colorlinks=true, linkcolor=red, citecolor=blue]{hyperref}
\usepackage{longtable}

\begin{document}

\title{Cosmic Constraints to $w$CDM Model from Strong Gravitational Lensing}

\author{Jie An}
\author{Baorong Chang}
\author{Lixin Xu}
\email{lxxu@dlut.edu.cn}

\affiliation{Institute of Theoretical Physics, School of Physics and Optoelectronic Technology, Dalian University of Technology, Dalian, 116024, People's Republic of China}

\begin{abstract}
In this paper, we study the cosmic constraint to $w$CDM model via $118$ strong gravitational lensing systems which are complied from SLACS, BELLS, LSD and SL2S surveys, where the ratio between two angular diameter distances $D^{obs} = D_A(z_l,z_s)/D_A(0,z_s)$ is taken as a cosmic observable. To obtain this ratio, we adopt two strong lensing models: one is the singular isothermal sphere model (SIS), the other one is the power-law density profile (PLP) model. Via the Markov Chain Mote Carlo method, the posterior distribution of the cosmological model parameters space is obtained. The results show that the cosmological model parameters are not sensitive to the parameterized forms of the power-law index $\gamma$. Furthermore, the PLP model gives a relative tighter constraint to the cosmological parameters than that of the SIS model. The predicted value of $\Omega_m=0.31^{+0.44}_{-0.24}$ by SIS model is compatible with that obtained by {\it Planck}2015: $\Omega_{m}=0.313\pm0.013$. However, the value of $\Omega_m=0.15^{+0.13}_{-0.11}$ based on the PLP model is smaller and has $1.25\sigma$ tension with that obtained by {\it Planck}2015 result. This discrepancy maybe come from the systematic errors.

\end{abstract}

\maketitle

\section{Introduction}

Dark energy is a synonym for a formalism which can describe the late time accelerated expansion of our Universe \cite{ref:AG1998,ref:SP1999}. Up to now, we have already known that a dark energy has a (effective) negative pressure (or a negative equation of state) and permeates the whole Universe smoothly, please see \cite{ref:book} and references therein for reviews. Revealing the properties of dark energy strongly depends on the cosmic observations. Since the property of dark energy mainly comes into the expansion rate of our Universe, i.e. the Hubble parameter $H(z)$, the geometrical measurements from the luminosity distance of type Ia supernovae and the sound horizon of baryon acoustic oscillations can provide a constraint to the equation of state of dark energy. In the literature, to grasp the main properties and to avoid building a concrete dark energy model, for instance selecting a concrete potential form $V(\phi)$ for a quintessence dark energy model, one chooses a parameterized equation of state of dark energy, i.e. $p=w\rho$. This $w$CDM model can be taken as the simplest extension to the $\Lambda$CDM model where $w=-1$ is respected. Now the dark energy issue is how to determine the values of $w$ at different redshift $z$ from cosmic observations. Since this simplest parameterization of $w$ connects to the cosmic observations through the Friedmman equation or the Hubble parameter $H(z)$ as a function of redshift $z$, the geometric measurements, such as the luminosity distance of type Ia supernovae \cite{ref:AG1998,ref:SP1999,ref:Riess2004,ref:Rosati2002,ref:Kowalski2008}, the position of the peaks of the cosmic microwave background (CMB) radiation \cite{ref:Komatsu2009,ref:Spergel2007} and the baryon acoustic oscillations \cite{ref:Eisenstein2005} and gamma-ray bursts \cite{ref:XuGSN,ref:Xugamma,ref:Xuchapgamma,ref:XuGRB,ref:Wang2015} can constrain the values of $w$. Principally, more cosmic observations are needed to get a tighter constraint to the properties of dark energy, here its equation of state.

When the light rays pass through astronomical objects (galaxies, cluster of galaxies), the cosmological gravitation field bends the paths traveled by light from distant source to us. These light paths respond to the distribution of mass. Therefore, the positions of the source and the image are related through the simple lens equation. The information of the invisible matter can be deduced by making a thorough inquiry the relation between sources and images. If one can measure the redshifts of the source and lens, the velocity dispersion of the mass distribution, the separated image, then one might be able to infer the distribution of the mass in our Universe.  Since the first discovery of the strong gravitational lensing in $Q0957+561$ by Walsh, Carswell and Weymann in 1979 \cite{ref:Walsh1979}, it has become a powerful probe in the study of cosmology. Actually, in the last few years, the strong lensing systems have been used as useful probe to determine the cosmological model parameter space \cite{ref:Biesiada2006,ref:Biesiada2010,ref:Cao2012,ref:wangnan2013,Zhu,hong,Chae,Mitchell,ref:Treu2006,ref:Koopmans2009,ref:Kochanek1996,ref:Ofek2003,ref:Chae2003,ref:Chae2004,ref:chenyun2015,ref:Cao2015,ref:Li2015,ref:Liao2015}. Up to now, the data points have amounted to $118$ as collected in Ref. \cite{ref:Cao2015}. These $118$ strong lensing systems are complied from four surveys: SLACS, BELLS, LSD and SL2S. The Sloan Lens ACS Survey (SLACS) and the BOSS emission-line lens survey (BELLS) are spectroscopic lens surveys in which candidates are selected respectively from Sloan Digital Sky Survey (SDSS) data and Baryon Oscillation Spectroscopic Survey (BOSS). BOSS has been initiated by upgrading SDSS-I optical spectrograph \cite{ref:Eisenstein2011}. These data points allow us to investigate the properties of dark energy.

Strong gravitational lensing occurs whenever the source, the lens and the observer are so well aligned that the observer-source direction lies inside the so-called Einstein ring of the lensing. To extract the geometric information from a strong lensing system, one should specify a lensing model connecting the source, the image. In fact, the geometric information can be derived is the ratio
\begin{equation}
D^{obs}(z_l,z_s)=D_{A}(z_l,z_{s})/D_{A}(0,z_{s}),
\end{equation}
between two angular diameter distances $D_{A}(z_l,z_{s})$ (the angular diameter distance between lens and source) and $D_{A}(0,z_{s})$ (the angular diameter distance between the observer and the source) which has been adopted as an observable \cite{ref:Biesiada2006,ref:Biesiada2010,ref:Cao2012,ref:wangnan2013}. In theory, this angular diameter distance reads
\begin{equation}
D_A(z_1,z_2;p) = \frac{c}{H_0}\frac{1}{1+z_2}\int_{z_1}^{z_2}\frac{dz'}{E(z';p)},\quad (z_2>z_1),
\end{equation}
for a spatially flat cosmology, where $H_0$ is the Hubble constant and $E(z;p)$ is the dimensionless expansion
rate which depends on redshift $z$ and cosmological parameters $p$. For the $w$CDM model considered in this paper, $p$ should be
 \begin{equation}
p=\{H_0, \Omega_m,w\},
\end{equation}
and the dimensionless expansion rate reads
\begin{equation}
E^2(z;p) = \Omega_m (1+z)^3+\Omega_w(1+z)^{3(1+w)},
\label{eq:E2z}
\end{equation}
where $\Omega_m$ is dimensionless density parameter of matter, and $\Omega_w=1-\Omega_m$ is dimensionless density parameter of dark energy for a spatially flat cosmology. It is clear that the Hubble parameter $H_0$ cannot be constrained, once the ratio $D^{obs}(z_l,z_s)=D_{A}(z_l,z_{s})/D_{A}(0,z_{s})$ is adopted as cosmic observable. Now we have only two cosmological model parameters, $ \Omega_m,w$ remained to be determined by the strong lensing systems. Of course, the free model parameter would be increased for a specific lensing model.

Recently, the authors of Ref. \cite{ref:Cao2015} studied the equation of state of dark energy by using $118$ strong lensing systems under the assumption of power-law density profile. In Ref. \cite{ref:Cao2015}, it also claimed that the strong lensing systems cannot give a tight constraint to the density parameter of dark matter $\Omega_m$ due to the possible narrow range of values of the ratio $D^{obs}(z_l,z_s)=D_{A}(z_l,z_{s})/D_{A}(0,z_{s})$. Therefore, $\Omega_m$ was fixed to the best-fit value obtained by {\it Planck}2013 in Ref. \cite{ref:Cao2015}. It is clear that the results are limited. Since the number of the strong lensing systems amounts to $118$ which is almost twice as much as that in before, one would like to see the constraining power for $\Omega_m$. On the other hand, once the power-law density profile is assumed, one would like to understand the degeneracy between $\Omega_m$ and the power law index, and also the impact on the cosmological parameters due to the evolution of the power law index. This is the main motivation of this work.

This paper is organized as follows: In Section \ref{sec:method}, we simply introduce strong gravitational lensing models: the SIS model and power-law density profile model respectively. In Section \ref{sec:results}, we
will present constraint results to $w$CDM model using the SIS model and power-law density profile models by relaxing the values of $\Omega_m$. We also study the impact on the cosmological parameters due to the different parameterization of $\gamma(z_l)$ and its degeneracy to $\Omega_m$ in this section. Conclusion is given in Section \ref{sec:conclusion}.

\section{Lensing Models} \label{sec:method}

To have the ratio of $D^{obs}(z_l,z_s)=D_{A}(z_l,z_{s})/D_{A}(0,z_{s})$ for a strong lensing system, one needs to specify a lensing model which relates the Einstein radius to the projected mass density. In this paper, we will mainly focus on the
singular isothermal sphere (SIS) or singular isothermal ellipsoid (SIE) profile \cite{ref:Treu2006,ref:Koopmans2009,ref:Kochanek1996} and a generalized spherically symmetric power-law mass distribution model $\rho \sim r^{-\gamma}$. The spherically symmetric power-law model was proposed to consider the possible deviation from the isothermal profile and its evolution with redshift due to the structure formation theory \cite{ref:Ruff2011,ref:Brownstein2012,ref:Sonnenfeld}. The singular isothermal sphere (SIS) is recovered when $\gamma = 2$ is respected. In the following part of this section, we will give a brief review of these two models.

\subsection{Singular Isothermal Sphere Model}

Analysis of isothermal mass properties have been widely applied in statistical analysis of the lens \cite{ref:Treu2006,ref:Koopmans2009,ref:Kochanek1996}. Under the first order approximation, SIS model is a good choice to obtain mean parameters of galaxies in the statistical analysis. The Einstein radius in SIS model is given by \cite{ref:Ofek2003,ref:Cao2012}
\begin{equation}
\theta_{E}=4\pi\frac{D_{A}(z_l,z_{s})}{D_{A}(0,z_{s})}\frac{\sigma_{SIS}^2}{c^2},\label{eq:ER}
\end{equation}
where $\sigma_{SIS}$ is the velocity dispersion and $c$ is the speed of light. One introduces the relation between the stellar velocity dispersion $\sigma_0$ and  the velocity dispersion $\sigma_{SIS}$ in the form of
\begin{equation}
\sigma_{SIS}=f_{E}\sigma_0,\label{eq:relation}
\end{equation}
where $f_E$ is a free model parameter. Then if the observed values of the Einstein radius and the velocity dispersion are obtained, one has the observed ratio easily
\begin{equation}
D^{obs}_{SIS} =\frac{ D_A(z_l, z_s)}{D_A(0, z_s)}= \frac{c^{2}\theta_E}{4\pi {f_{E}}^2 {\sigma_0}^2},\label{eq:SISobs}
\end{equation}
here $\theta_E$ and $\sigma_0$ are the Einstein radius and the stellar velocity dispersion respectively; $c$ is the speed of light. According to \cite{ref:Ofek2003,ref:Cao2012}, the value of $f_E$ is in the range $0.8^{1/2}<f_{E}<1.2^{1/2}$. In order to eliminate the effects and uncertainties caused by the free parameter $f_E$, Wang and one of us proposed to use the ratio $\mathcal{D}^{obs}_{ij}=\theta_{E_{i}}\sigma_{0_{j}}^2/\theta_{E_{j}}\sigma_{0_{i}}^2$ as cosmic observations to constrain the cosmological model, where $i$, $j$ denote the order numbers of the lensing systems used in Ref. \cite{ref:wangnan2013}. In this way, the effects and uncertainties caused by the free
parameter $f_E$ are eliminated completely. But in this paper, $f_E$ together with the cosmological model parameters will be treated as a free model parameters.

When we obtain the observable $D^{obs}$ from the observed values of the Einstein radius $\theta_E$ and the velocity dispersion $\sigma_0$, the uncertainties are also introduced into the error budget. Via the error propagation equation, one can derive the $1\sigma$ error for $D^{obs}_{SIS}$ via the equation
\begin{equation}
\sigma^2_{D(SIS)} = \left(D^{obs}_{SIS}\right)^2\left[\left(\frac{\sigma_{\theta_E}}{\theta_E}\right)^2+4\left(\frac{\sigma_{\sigma_0}}{\sigma_0}\right)^2+4\left(\frac{\sigma_{f_E}}{f_E}\right)^2\right],
\label{eq:sigmaSIS}
\end{equation}
where $\sigma_{\theta_E}$, $\sigma_{\sigma_0}$ and $\sigma_{f_E}$ are the uncerteinties of $\theta_E$, $\sigma_0$ and $f_E$ respectively. Fortunately, $\sigma_0$ and $\sigma_{\sigma_0}$ can be obtained from observations. Following the SLACS team, we take $5\%$ error for $\theta_E$ and $f_E$ due to the fractional uncertainty of the Einstein radius at the level of $5\%$.  According to Ref. \cite{ref:Grillo2008}, it can be seen that choose $\sigma_{\theta_E}$ the level of $5\%$ is reasonable, see also in Ref. \cite{ref:Cao2015}.

\subsection{Power-law Density Profile Model}

For a general power-law density profile model, the mass density distribution is given by $\rho\sim r^{-\gamma}$ according to Refs. \cite{ref:Cao2015}.  Since lensing mass inside the Einstein radius is determined by $\theta_E$, for a spherically symmetric lensing system, we can shift the coordinate origin to the centre of symmetry and reduce light
deflection to a one-dimensional problem,
\begin{equation}
M(\xi) = 2\pi\int^{\xi}_{0}\Sigma(\xi')\xi'd\xi',
\end{equation}
where $M(\xi)$ is the mass enclosed within the radius $\xi$, and $\Sigma(\xi)$ is the projected mass density for lensing. For the case of a Einstein ring, the mass $M_{lens}$ inside Einstein radius is given by
\begin{equation}
M_{lens} = \pi\Sigma_{cr}R_E^2,
\end{equation}
where $R_E = \theta_E D_A(z_l) $ is the physical Einstein radius, and $\Sigma_{cr}$ is the critical projected mass density
\begin{equation}
\Sigma_{cr} = \frac{c^{2}}{4\pi G}\frac{D_A(0,z_s) }{D_A(0,z_l)D_A(z_l,z_s)}.
\end{equation}
Thus the lens mass $M_{lens}$ inside the Einstein radius reads
\begin{equation}
M_{lens} = \frac{c^2}{4G}\frac{D_A(0,z_s)D_A(0,z_l)}{D_A(z_l,z_s)}\theta^2_E.
\label{MLENS}
\end{equation}
After solving spherical Jeans equation, one can assess the dynamical mass inside the aperture projected to lens plane and scale it to the Einstein radius
\begin{equation}
M_{dyn} = \frac{\pi}{G}\sigma^2_{ap}D_A(0,z_l)\theta_E\left(\frac{\theta_E}{\theta_{ap}}\right)^{2-\gamma}f(\gamma),
\label{Mdyn}
\end{equation}
where
\begin{equation}
f(\gamma) = -\frac{1}{\sqrt{\pi}}\frac{(5-2\gamma)(1-\gamma)}{3-\gamma} \frac{\Gamma(\gamma-1)}{\Gamma(\gamma-3/2)} \left[\frac{\Gamma(\gamma/2-1/2)}{\Gamma(\gamma/2)}\right]^2,
\end{equation}
and $\sigma_{ap}$ is the velocity dispersion inside the aperture. The relationship between $\sigma_{ap}$ and $\sigma_0$ is given by \cite{ref:IJ1995a,ref:IJ1995b}
\begin{equation}
\sigma_0 = \sigma_{ap}\left(\frac{\theta_{eff}}{2\theta_{ap}}\right)^{0.04}.
\end{equation}
By setting $M_{lens} =M_{dyn}$, one obtains $\theta_E$
\begin{equation}
\theta_{E} = 4\pi \frac{\sigma^2_{ap}}{c^2}\frac{D_A(z_l,z_s)}{D_A(0,z_s)}\left(\frac{\theta_E}{\theta_{ap}}\right)^{2-\gamma}f(\gamma).
\end{equation}
Now we have the observable in the power-law density profile model
\begin{equation}
D^{obs}_{PLP} = \frac{c^2\theta_E}{4\pi \sigma^2_{ap}}\left(\frac{\theta_{ap}}{\theta_{E}}\right)^{2-\gamma}f^{-1}(\gamma).\label{eq:PLPobs}
\end{equation}
In this model, the $1\sigma$ error of $D_{PLP}^{obs}$ comes from the uncertainties of $\sigma_{ap}$ and $\theta_E$ following Ref. \cite{ref:Cao2015}
\begin{equation}
\sigma^2_{D(PLP)} = \left(D^{obs}_{PLP}\right)^2 \left[4\left(\frac{\sigma_{\sigma_{ap}}}{\sigma_{ap}}\right)^2+\left(1-\gamma\right)^2\left(\frac{\sigma_{\theta_E}}{\theta_E}\right)^2\right],
\label{eq:sigmaPL}
\end{equation}
here we also take $5\%$ error for $\theta_E$, and the variance of $\sigma_{ap}$ has presented in Ref. \cite{ref:Cao2015}.

It was suggested that the mass density power-law index $\gamma$ of massive elliptical galaxies evolves with respect to redshift \cite{ref:Ruff2011,ref:Brownstein2012,ref:Sonnenfeld}. And a linear relation with $z_l$ was assumed $\gamma(z_l)=2.12^{+0.03}_{-0.04}-0.25^{+0.10}_{-0.12}\times z_l+0.17\pm 0.02 (\text{scatter})$ by combining the lensing sample from SLACS, SL2S and LSD. Here we just quote the results in the case of $w$CDM model obtained in Ref. \cite{ref:Cao2015}: $\gamma_0=2.06\pm0.09$, $\gamma_1=-0.09\pm0.16$ where $118$ lensing systems were used under the assumption of a fixed  best-fit value of $\Omega_m$ obtained by {\it Planck}2013. Although, by this assumption, a relative tighter constraint to the equation of sate of dark energy could be obtained as shown in Ref. \cite{ref:Cao2015}, the results and conclusion are limited to the special case. More importantly, one would like to see the power in constraining cosmological model beyond the only equation of sate of dark energy, say $\Omega_m$. Furthermore, we would like to know the possible degeneracy between $\Omega_m$ and $\gamma$, and the possible dependence of cosmological model parameter space on the parameterzied forms of $\gamma$. Therefore, in this work, we choose two parameterized forms of $\gamma$:
\begin{eqnarray}
\rm{Model~I:}\quad \gamma(z_l) &=& \gamma_0 + \gamma_a z_{l},\\
\rm{Model~II:}\quad \gamma(z_l) &=& \gamma_0 + \gamma_a \frac{z_{l}}{1+z_{l}},
\end{eqnarray}
to study the parameterzied form dependence issue, where $\gamma_0$ and $\gamma_a$ are free constant parameters. The Model II is inspired by the CPL dark energy model which was already taken as a very nature Taylor expansion at present, i.e. $a=1$. As a comparison to Model I, Model II is not divergent even at quite large redshift, although the lens distribute at relative low redshift. To see the constraining power to the cosmological model and the possible degeneracy between $\Omega_m$ and $\gamma$, we relax $\Omega_m$ as a free cosmological parameter.

\section{Methodology and Results} \label{sec:results}

To obtain the model parameter space, we perform a global fitting via the Markov Chain Monte Carlo method which is based on the publicly available {\bf CosmoMC} package \cite{ref:AL2002}. The posterior likelihood $\mathcal{L}\sim\exp\left[-\chi^2(p)/2\right]$ is given by calculating the $\chi^2(p)$
\begin{equation}
\chi^2(p)= \sum_{i=1}^{118}\frac{(D_{i}^{th}(p)-D_{i}^{obs})^2}{\sigma_{D,i}^2},
\label{eq:chi2}
\end{equation}
where $p$ is the model parameter vector, $i$ denotes the $i$th strong lensing system, $\sigma_{D,i}^2$ is the corresponding $1\sigma$ variance of $D^{obs}_{i}$. The theoretical values of $D_{i}^{th}(p)$ is calculated by
\begin{equation}
D_{i}^{th}(p)=\frac{D_{A}(z_l,z_{s};p)}{D_{A}(0,z_{s};p)}=\frac{\int_{z_l}^{z_s}\frac{dz'}{E(z';p)}}{\int_{0}^{z_s}\frac{dz'}{E(z';p)}},
\end{equation}
where $z_l$ and $z_s$ are the redshifts at the lens and source of the lensing system respectively. For the SIS (and PLP) model, the observed value $D_{i}^{obs}$ is derived from Eq. (\ref{eq:SISobs}) (and Eq. (\ref{eq:PLPobs})) and its $1\sigma$ error $\sigma_{D,i}$ is calculated by Eq. (\ref{eq:sigmaSIS}) (and Eq. (\ref{eq:sigmaPL})). The  corresponding redshifts of the $118$ lensing systems, the observed values of $\theta$s, $\sigma$s and their $1\sigma$ error bars can be found in Table I of Ref. \cite{ref:Cao2015}.

\subsection{Constraint Results for the SIS Model}

\begin{center}
\begin{figure}[!htbp]
\includegraphics[width=9cm]{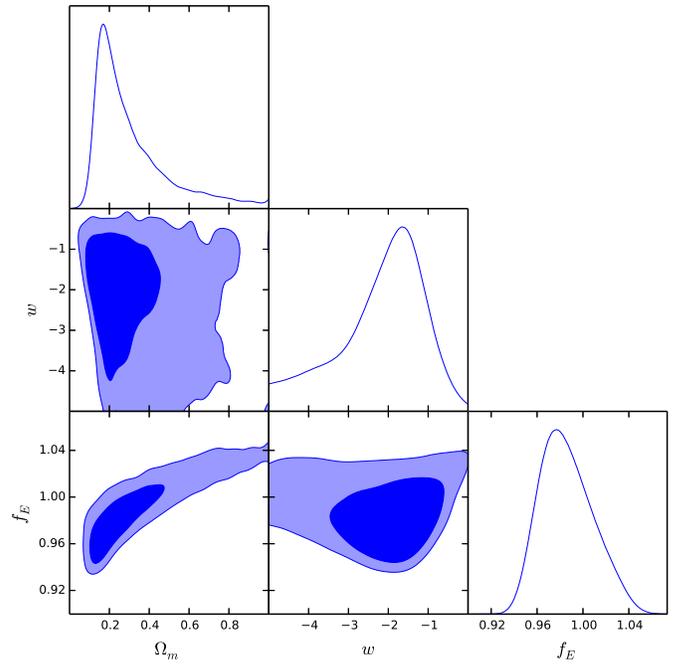}
\caption{The $68.3\%$ and $95.4\%$ confidence regions for $w$CDM model obtained from 118 strong lensing systems in SIS model.}\label{fig:SIS}
\end{figure}
\end{center}

In the SIS model, we have three cosmological model parameters and one SIS model parameter, i.e.
\begin{equation}
p=\{H_0, \Omega_m, w, f_{E}\}.
\end{equation}
Because the ratio $D_{A}(z,z_{s})/D_{A}(0,z_{s})$ is taken as observable, the Hubble parameter $H_0$ cannot be constrained, therefore $H_0$ is marginalized in our analysis in SIS model and power-law density profile models. The constraint results are shown in Table \ref{tab:SIS} and Figure \ref{fig:SIS}. In $1\sigma$ region, one has $\Omega_m=0.31^{+0.44}_{-0.24}$ which is compatible with that obtained by {\it Planck}2015 result: $\Omega_{m}=0.313\pm0.013$ \cite{ref:Planck2015}. The value of $w=-2.2^{+1.7}_{-2.4}$ in $1\sigma$ region implies that our Universe is undergoing an accelerated expansion. It confirms the findings by SN Ia independently. The value of $f_E=0.986^{+0.047}_{-0.041}$ centering around $f_E\sim 1$ is also compatible with previous results \cite{ref:Ofek2003,ref:Cao2012}.
\begin{center}
\begin{table}
\begin{tabular} {cc}
\hline\hline
 Parameter &  95\% limits \\ \hline
$\Omega_m$ & $0.31^{+0.44}_{-0.24}      $\\
$w$ & $-2.2^{+1.7}_{-2.4}$\\
$f_E$ & $0.986^{+0.047}_{-0.041}   $\\
\hline\hline
\end{tabular}
\caption{For SIS model: the mean values with $1\sigma$ errors of the parameters for the $w$CDM model, where the combined
observational data of 118 strong lensing systems from SLACS, BELLS, LSD and SL$2$S.} \label{tab:SIS}
\end{table}
\end{center}

\subsection{Constraint Results for the Power-law Density Profile Model}

\begin{center}
\begin{figure}[!htbp]
\includegraphics[width=9cm]{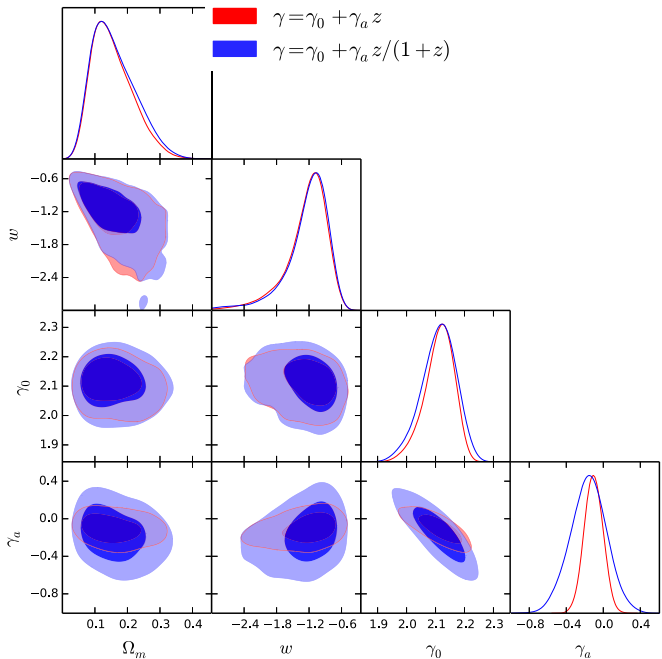}
 \caption{The $68.3\%$ and $95.4\%$ confidence regions for $w$CDM model obtained from 118 strong lensing systems for the power-law density profile models: Model I $\gamma(z_l) = \gamma_0 + \gamma_a z_{l}$ and Model II $\gamma(z_l) = \gamma_0 + \gamma_a z_{l}/(1+z_{l})$.}\label{fig:PL}
\end{figure}
\end{center}

In this model, we also have three cosmological model parameters and two profile index parameter, i.e.
\begin{equation}
p=\{H_0, \Omega_m, w, \gamma_0, \gamma_a\}.
\end{equation}
As stated in the previous subsection, $H_0$ is marginalized. The constraint results are shown in Table \ref{tab:PL} and Figure \ref{fig:PL}. As shown in Figure \ref{fig:PL}, we obtain almost the same distribution for the cosmological model space, i.e. $\Omega_{m}=0.15^{+0.13}_{-0.11}$, $w=-1.25^{+0.66}_{-0.86}$ for the Model I and $\Omega_m=0.16^{+0.14}_{-0.11}$ and $w=-1.23^{+0.66}_{-0.87}$ for the Model II, although the distribution for $\gamma$ parameters are different. It means that the cosmological model parameters, $\Omega_m$ and $w$, are not sensitive to the parameterization form of $\gamma$. And there is no significant difference for the $\Delta\chi^2=0.7$ for two parameterized forms. Therefore, the cosmological parameter space do not depend on the parameterized forms of $\gamma$. And in the contour plot for $\Omega_m-\gamma$, there is no significant degeneracy between $\Omega_m$ and $\gamma$. But $\gamma_0$ and $\gamma_a$ are anti-correlated for centring around $\gamma \sim 2$.

As a comparison to the SIS model, the power-law density profile model can give a relative tight constraint to the cosmological model parameters and favors small values of $\Omega_m$. But we should note that the values of $\Omega_m=0.15^{+0.13}_{-0.11}$ have $1.25\sigma$ tension with that obtained by {\it Planck}2015 result: $\Omega_{m}=0.313\pm0.013$ \cite{ref:Planck2015}.
\begin{center}
\begin{table}
\begin{tabular} {ccc}
\hline\hline
 Parameter &  $95\%$ limits Model I & $95\%$ limits Model II  \\ \hline
$\Omega_m$ & $0.15^{+0.13}_{-0.11}     $ & $0.16^{+0.14}_{-0.11}      $\\
$w$ & $-1.25^{+0.66}_{-0.86}    $ & $-1.23^{+0.66}_{-0.87}     $\\
$\gamma_0$ & $2.11^{+0.10}_{-0.11}   $& $2.11^{+0.11}_{-0.12}      $\\
$\gamma_a$ & $-0.10^{+0.19}_{-0.20}    $ & $-0.16^{+0.39}_{-0.40}     $\\
\hline\hline
\end{tabular}
\caption{The same as SIS model but for the power-law density profile models: Model I $\gamma(z_l) = \gamma_0 + \gamma_a z_{l}$ and Model II $\gamma(z_l) = \gamma_0 + \gamma_a z_{l}/(1+z_{l})$.}\label{tab:PL}
\end{table}
\end{center}

\section{Conclusion}  \label{sec:conclusion}

In this paper, we show the constraint to the $w$CDM model  through $118$ strong lensing systems by relaxing $\Omega_m$ as a free cosmological parameter which was previously fixed to its best-fit values obtained by {\it Planck}2013 in Ref. \cite{ref:Cao2015}. Our results are obtained based on two lensing models which are characterized by its mass density profile, i.e. the singular isothermal sphere (SIS) model and the power-law density profile (PLP) model. For PLP model, to study the possible degeneracy between $\Omega_m$ and $\gamma$, and the possible dependence of cosmological model parameter space on the parameterzied forms of $\gamma$, we proposed two parameterized forms for the density power index $\gamma$: $\gamma=\gamma_0+\gamma_a z_l$ (Model I) and $\gamma=\gamma_0+\gamma_a z_l/(1+z_l)$. Via the Markov Chain Monte Carlo method, the posterior distribution of the cosmological parameters space and lensing model parameter space are obtained as shown in Table \ref{tab:SIS} and Table \ref{tab:PL}. For the SIS model, one obtains $\Omega_m=0.31^{+0.44}_{-0.24}$ which is compatible with that obtained by {\it Planck}2015 result: $\Omega_{m}=0.313\pm0.013$ \cite{ref:Planck2015} but large $1\sigma$ error bars. And the values of $w=-2.2^{+1.7}_{-2.4}$ in $1\sigma$ region imply that our universe is undergoing an accelerated expansion. It confirms the findings by SN Ia independently. For the PLP model, as a comparison to the SIS model, a relative tight constraint is obtained for the cosmological parameter space: $\Omega_m=0.15^{+0.13}_{-0.11}$, $w=1.25^{+0.66}_{-0.86}$. But PLP model favours small values of $\Omega_m$ which also has $1.25\sigma$ tension with that of {\it Planck}2015 result. We should notice that the systematic errors of lensing systems are not included, this discrepancy maybe come from the systematic errors. For this point, it still under investigation. We do not see any significant dependence on the parameterized form of $\gamma$ for the cosmological parameter space. But $\gamma_0$ and $\gamma_a$ is anti-correlated for keeping $\gamma$ around $\gamma \sim 2$.

\acknowledgements{We thank an anonymous referee for helpful improvement of this paper. L. Xu's work is supported in part by NSFC under the Grants No. 11275035.}


\begin{thebibliography}{*}
\bibitem{ref:AG1998} A. G. Riess, et al., Astrophys. J. 116, 1009(1998).
\bibitem{ref:SP1999} S. Perlmutter, et al., Astrophys. J. 517, 565(1999).
\bibitem{ref:book} L. Amendola, S. Tsujikawa, Cambridge University Press, Dark Energy: Theory and Observations(2010).
\bibitem{ref:Riess2004} A. G. Riess, et al., Astrophys. J. 607, 665(2004).
\bibitem{ref:Rosati2002} P. Rosati, S. Borgani, C. Norman, Astrono.\& Astrophy.  40, 539(2002).
\bibitem{ref:Kowalski2008} M. Kowalski, et al., Astrophys. J. 686, 749(2008).
\bibitem{ref:Komatsu2009} E. Komatsu, et al.,Astrophys. J. 180, 330(2009).
\bibitem{ref:Spergel2007} D. N. Spergel, et al., Astrophys. J. Suppl. Ser. 170, 377(2007).
\bibitem{ref:Eisenstein2005} D. J. Eisenstein, et al., Astrophys. J. Suppl. Ser. 633, 560(2005).

\bibitem{ref:XuGSN} L. Xu, Y. Wang, J. Cosmol. Astropart.
    Phys. 11, 104(2010), arXiv:1007.4734 [astro-ph.CO].
\bibitem{ref:Xugamma} L. Xu, Y. Wang, Phys. Lett. B, 702, 114(2011), arXiv:1009.0963 [astro-ph.CO].
\bibitem{ref:Xuchapgamma} N. Liang, L. Xu, Z. H. Zhu, Astrono.\& Astrophy. 527, A11 (2011), arXiv:1009.6059 [astro-ph.CO].
\bibitem{ref:XuGRB} L. Xu,  J. Cosmol. Astropart.
    Phys. 04, 025(2012), arXiv:1005.5055 [astro-ph.CO].
\bibitem{ref:Wang2015} F. Y. Wang, Z. G. Dai, E. W. Liang, New Astr. Rev. 67, 1(2005).

\bibitem{ref:Walsh1979}  D. Walsh, R. F. Carswell, R. J. Weymann, Nature. 279, 381(1979).
\bibitem{ref:Eisenstein2011} D. J. Eisenstein, et al., Astrophys. J. 142, 72(2011).
\bibitem{ref:Biesiada2006} M. Biesiada, Phys. Rev. D. 73, 023006(2006).
\bibitem{ref:Cao2015} S. Cao, M. Biesiada, R. Gavazzi, A. Piorkowska, Z. H. Zhu, Astrophys. J. 806, 185(2015).
\bibitem{ref:Biesiada2010} M. Biesiada, A. Piorkowska, and B. Malec, Mon. Not. R. Astron. Soc. 1059(2010).
\bibitem{ref:Cao2012} S. Cao and Z. H. Zhu, J. Cosmol. Astropart.
    Phys. 03, 016 (2012) [astor-ph/1105.6226].
\bibitem{ref:wangnan2013} N. Wang and L. Xu, Mod. Phys. Lett. A 28, 1350057 (2013).
\bibitem{Zhu} Z. H. Zhu, Mod. Phys. Lett. A 15, 1023 (2000) [astor-ph/0010351].
\bibitem{hong} Z. H. Zhu, Int. J. Mod. Phys. D 9, 591 (2000) [astor-ph/0107233].
\bibitem{Chae} K. H. Chae, Mon. Not. Roy. Asrron. Soc. 346, 746 (2003).
\bibitem{Mitchell} J. L. Mitchell et al., Astrophys. J. 622, 81 (2005) [astor-ph/0401138].
\bibitem{ref:Treu2006} T. Treu, L V Koopmans and A S Bolton, et al, Astrophys. J. 640: 662-672.(2006)[astroph/0512044].
\bibitem{ref:Koopmans2009} L V E. Koopmans, A Bolton and T Ture, et al, Astrophys. J. 115:  377.(2009).
\bibitem{ref:Kochanek1996} C S. Kochanek, Astrophys. J. 466:  638-659.(1996).
\bibitem{ref:Ruff2011} A. Ruff, et al., Astrophys. J. 727, 96(2011).
\bibitem{ref:Brownstein2012} Brownstein, et al., Astrophys. J. 744, 41(2012)
\bibitem{ref:Sonnenfeld} A. Sonnenfeld, et al., Astrophys. J. 777, 98(2013), arXiv:1307.4759 [astro-ph.CO].

\bibitem{ref:Ofek2003} E. O. Ofek, H. W. Rix and D. Maoz, Mon. Not. R. Astron. Soc. 343, 639(2003).

\bibitem{ref:chenyun2015} Y. Chen, C. Geng, S. Cao, Y. Huang, and Z. Zhu, [astro-ph/04011384](2015).
\bibitem{ref:Chae2003} K. H. Chae, S. Mao, Astrophys. J. 599, L61(2003).
\bibitem{ref:Chae2004} K. H. Chae et al., Astrophys. J. 607, L71(2004).

\bibitem{ref:Li2015} X. Li, S. Cao, X. Zheng, M. Biesiada, Z. H. Zhu, arXiv:1510.03494 [astro-ph.CO].
\bibitem{ref:Liao2015} K. Liao, Z. Li, S. Cao, M. Biesiada, X. Zheng, Z. H. Zhu, arXiv:1511.01318 [astro-ph.CO].
\bibitem{ref:IJ1995a} I. J$\phi$rgensen, M. Franx and P. Kj{\ae}rgard. Mon. Not. R. Astron. Soc. 273, 1097(1995).

\bibitem{ref:IJ1995b} I. J$\phi$rgensen, M. Franx and P. Kj{\ae}rgard. Mon. Not. R. Astron. Soc. 276, 1341(1995).

\bibitem{ref:Grillo2008} C. Grillo and M. Lombardi et al, Astrono.\& Astrophy, 477 397(2008).
\bibitem{ref:AL2002} A. Lewis and S. Bridle, Phys. Rev. D. 66 103511(2002).
\bibitem{ref:Planck2015} [Planck Collaboration]: P. A. R. Ade, et al., arXiv:1502.01589v2 [astro-ph.CO].


\end{thebibliography}
\end{document}